\documentclass[11pt]{article}

\newcommand{\be}{\begin{equation}}
\newcommand{\ee}{\end{equation}}
\newcommand{\bi}[1]{\vspace{-3mm} \bibitem{#1}}

\begin{document}

\begin{center}
Modern Physics Letters B. Vol.19. No.22. (2005) pp.1107-1118. 
\end{center}

\begin{center}
{\Large \bf Multipole Moments of Fractal Distribution of Charges}
\vskip 5 mm

{\large \bf Vasily E. Tarasov}\\

\vskip 3mm
{\it Skobeltsyn Institute of Nuclear Physics, \\
Moscow State University, Moscow 119992, Russia } \\
\end{center}
\vskip 11 mm

\noindent
In this paper we consider the electric multipole moments 
of fractal distribution of charges.
To describe fractal distribution, we use the fractional integrals.
The fractional integrals are considered as approximations of
integrals on fractals.  
In the paper we compute the electric multipole moments
for homogeneous fractal distribution of charges. \\



\section{Introduction}

Integrals and derivatives of fractional order have found 
many applications in recent studies in science.
The interest in fractals and fractional analysis has been growing 
continually in the last few years. 
Fractional derivatives and integrals have numerous applications: 
kinetic theories \cite{Zaslavsky1,Zaslavsky2,Physica};  
statistical mechanics \cite{chaos,PRE05,JPCS}; 
dynamics in complex media \cite{PLA05,AP05,Chaos05,PLA05-2,MPLB-1}; 
electrodynamics \cite{En1,En2,En3,En4} and many others. 
The new type of problem has increased rapidly in areas 
in which the fractal features of a process or the medium 
imposes the necessity of using non-traditional tools 
in smooth physical models. 
In order to use fractional derivatives and fractional integrals 
for fractal distribution, we must use some 
continuous medium model \cite{PLA05,AP05}. 
We propose to  describe the fractal distribution by a fractional 
continuous medium \cite{PLA05,AP05}, where all characteristics and 
fields are defined everywhere in the volume but they follow some 
generalized equations which are derived by using fractional integrals.
In many problems the real fractal structure of a medium
can be disregarded and the fractal distribution can be replaced by
some fractional continuous mathematical model.
By smoothing of microscopic characteristics over the
physically infinitesimal volume, we transform the initial
fractal distribution into a fractional continuous model \cite{PLA05,AP05}
that uses the fractional integrals.
The order of fractional integral is equal
to the fractal dimension of distribution.
The fractional integrals allow us to take into account the
fractality of the media \cite{PLA05}.
Fractional integrals are considered as approximations of 
integrals on fractals \cite{RLWQ}. 
It was proved that integrals
on net of fractals can be approximated by fractional integrals \cite{RLWQ}.
In Refs. \cite{chaos,PRE05}, we proved that fractional integrals
can be considered as integrals over the space with fractional
dimension up to a numerical factor. 

In this paper, we consider electric multipole moments  
of the fractal distribution of charges. 
Fractal distribution is described by the fractional continuous model
\cite{PLA05,AP05,Chaos05,PLA05-2}. 
In the general case, the fractal distribution cannot 
be considered as a continuous distribution.
There are domains that are not filled by particles.
We suggest \cite{PLA05} to consider the fractal distribution   
as a special (fractional) continuous distribution.
We use the procedure of replacement of the distribution 
with fractal dimension by some continuous model that 
uses fractional integrals. The suggested procedure can be considered as 
a fractional generalization of the Christensen approach \cite{Chr}.

In Sec. 2, the density of electric charge 
for fractal distribution is considered. 
In Sec. 3, we consider the electric multipole expansion.
In Sec. 4, the examples of electric dipole moment
for fractal distribution of chages are derived. 
In Sec. 5, we consider the electric quadrupole moment of
fractal distribution of charges.
In Sec. 6, the examples of electric quadrupole moments
of charged fractal parallelepiped are computed. 
In Sec. 7, the examples of electric quadrupole moments
of charged fractal ellipsoid are computed. 
Finally, a short conclusion is given in Sec. 8.

\section{Electric Charge of Fractal Distribution}

Let us consider a fractal distribution of charges.
For example, we can assume that charged particles 
with a constant density are distributed over the fractal. 
In this case, the number of particles $N(R)$ enclosed 
in a volume with characteristic size $R$ satisfies the scaling law 
\be  N(R) \sim R^{D} ,\ee 
whereas for a regular n-dimensional Euclidean object 
we have $N(R)\sim R^n$. 

For charged particles with number density $n({\bf r},t)$, we have that 
the charge density can be defined by
\be \rho({\bf r},t)=q n({\bf r},t) , \ee
where $q$ is the charge of particle 
(for example, q is an electron charge), and 
${\bf r}=x{\bf e}_1+y{\bf e}_2+z{\bf e}_3$. 
The total charge of region $W$ is then given by the integral
\be Q(W)=\int_W \rho({\bf r},t) dV_3= q\int_W n({\bf r},t) dV_3, \ee
i.e., $Q(W)=qN(W)$, where $N(W)$ is a number of particles in the region $W$. 
The fractional generalization of this equation can be written
in the following form:
\be Q(W)=\int_W \rho({\bf r},t) dV_D=q\int_W n({\bf r},t) dV_D , \ee
where $D$ is a fractal dimension of the distribution,
and $dV_D$ is an element of D-dimensional volume such that
\be \label{5a} dV_D=C_3(D,{\bf r})dV_3. \ee

For the Riesz definition of the fractional integral \cite{SKM}, the 
function $C_3(D,{\bf r})$ is defined by the relation
\be \label{5R} C_3(D,{\bf r})=
\frac{2^{3-D}\Gamma(3/2)}{\Gamma(D/2)} |{\bf r}|^{D-3} . \ee
The initial points of the fractional integral \cite{SKM} are set to zero.
The numerical factor in Eq. (\ref{5R}) has this form in order to
derive usual integral in the limit $D\rightarrow (3-0)$.
Note that the usual numerical factor
$\gamma^{-1}_3(D)=\Gamma(1/2)/(2^D \pi^{3/2} \Gamma(D/2))$,
which is used in Ref. \cite{SKM}
leads to $\gamma^{-1}_3(3-0)= \Gamma(1/2)/(2^3 \pi^{3/2}
\Gamma(3/2))=1/(4\pi^{3/2})$ in the limit $D\rightarrow (3-0)$.

For the Riemann-Liouville fractional integral \cite{SKM}, 
the function $C_3(D,{\bf r})$ is defined by
\be \label{5RL} C_3(D,{\bf r})=
\frac{|x y z |^{D/3-1}}{\Gamma^3(D/3)}  . \ee
Here we use Cartesian's coordinates $x$, $y$, and $z$. 
In order to have the usual dimensions of the physical values,
we can use vector ${\bf r}$, and coordinates 
$x$, $y$, $z$ as dimensionless values.

Note that the interpretation of fractional integration
is connected with noninteger dimension \cite{chaos,PRE05}.
This interpretation follows from the well-known formulas 
for dimensional regularizations.
The fractional integral can be considered as an
integral in the noninteger dimension space
up to the numerical factor $\Gamma(D/2) /( 2 \pi^{D/2} \Gamma(D))$.

If we consider the ball region $W=\{ {\bf r}: \ |{\bf r}|\le R \}$, 
Riesz fractional integral (\ref{5R}), 
and spherically symmetric distribution of charged particles 
($\rho({\bf r},t)=\rho(r)$), then we have
\be Q(W)=4\pi \frac{2^{3-D}\Gamma(3/2)}{\Gamma(D/2)}
\int^R_0 \rho(r) r^{D-1} dr . \ee
For the homogeneous ($\rho(r)=\rho_0$) fractal distribution, we get
\be Q(W)=\frac{4\pi \rho_0}{D} \frac{2^{3-D}\Gamma(3/2)}{\Gamma(D/2)}
R^D \sim R^D . \ee
If $D=3$, we have $Q(W)=(4 \pi/3) \rho_0 R^3$. 

If we consider the Riemann-Liouville fractional integral (\ref{5RL})
for the ball region $W$, and spherically symmetric distribution 
($\rho({\bf r},t)=\rho(r)$), then we have
\be Q(W)=\frac{2 \Gamma^3(D/6)}{\Gamma^3(D/3) \Gamma(D/2)}
\int^R_0 \rho(r) r^{D-1} dr . \ee
For the homogeneous ($\rho(r)=\rho_0$) fractal distribution, we get
\be Q(W)=\frac{2 \rho_0 \Gamma^3(D/6)}{D \Gamma^3(D/3) \Gamma(D/2)}
R^D \sim R^D . \ee
If $D=3$, we get the usual expression $Q(W)=(4 \pi/3) \rho_0 R^3$. 

Fractal distribution of charged particles is called a homogeneous 
fractal distribution if the power law  $Q(R)\sim R^D $ does not 
depend on the translation  of the region. 
The homogeneity property of the distribution 
can be formulated in the following form:
For all regions $W$ and $W^{\prime}$ such that the volumes 
are equal $V(W)=V(W^{\prime})$, 
we have that the number of particles in these regions 
are equal $N(W)=N(W^{\prime})$. 
Note that the wide class of the fractal media satisfies 
the homogeneous property.
In Refs. \cite{PLA05,AP05}, the continuous medium model for 
the fractal distribution was suggested.


\section{Electric Multipole Expansion for Fractal Distribution of Charges}
    
It is known that 
a multipole expansion is a series expansion of the effect produced 
by a given system in terms of an expansion parameter which becomes 
small as the distance away from the system increases. 
Therefore, the leading one of the terms in a multipole expansion are 
generally the strongest. The first-order behavior of the system 
at large distances can therefore be obtained from the first terms 
of this series, which is generally much easier to compute than 
the general solution. Multipole expansions are most commonly used 
in problems involving the electric and magnetic fields of charge and 
current distributions, and the propagation of electromagnetic waves.

To compute one particular case of a multipole expansion, 
let ${\bf R}=X_k{\bf e}_k$ be the vector from a fixed reference point to 
the observation point; ${\bf r}=x_k{\bf e}_k$ be the vector from 
the reference point to a point in the distribution; 
and ${\bf s}={\bf R}-{\bf r}$ be the vector from a point 
in the distribution to the observation point. 
The law of cosines then yields
\be \label{d} s^2=r^2+R^2-2rR \; \cos \theta \ee
where $s=|{\bf s}|$, $r=|{\bf r}|$, $R=|{\bf R}|$, and 
$\theta$ is the polar angle, defined such that
\be \cos \theta= ({\bf r},{\bf R})/(r R) . \ee
Using Eq. (\ref{d}), we get
\be s=R \sqrt{ 1 -2\frac{r}{R} \cos \theta +\frac{r^2}{R^2} } . \ee
Now define $\epsilon ={r}/{R}$, and $\xi=\cos \theta$, then
\be \label{1/d} \frac{1}{s}=\frac{1}{R} 
\Bigl( 1-2 \epsilon \xi+\epsilon^2 \Bigr)^{-1/2}. \ee
But the right hand side of Eq. (\ref{1/d})  
is the generating function for Legendre polynomials $P_n(\xi)$ 
by the following relation: 
\be \Bigl( 1-2 \epsilon \xi+\epsilon^2 \Bigr)^{-1/2}=
\sum^{\infty}_{n=0} \epsilon^n P_n(\xi) , \ee
so, we have the equation
\be \frac{1}{s}=\frac{1}{R} \sum^{\infty}_{n=0} 
\Bigl(\frac{r}{R}\Bigr)^n P_n( \cos \theta) . \ee
Any physical potential that obeys a $(1/s)$ law can therefore 
be expressed as a multipole expansion
\be \label{11}
V=  \frac{1}{4 \pi \varepsilon_0}  \sum^{\infty}_{n=0} \frac{1}{R^{n+1}}
\int_W r^n P_n( \cos \theta)  \rho({\bf r}) dV_D. 
\ee
The $n = 0$ term of this expansion, called the monopole term, 
can be pulled out by noting that $P_0(x)=1$, so
\be
V= \frac{1}{4 \pi \varepsilon_0} \frac{1}{R} \int_W \rho({\bf r}) dV_D+
\frac{1}{4 \pi \varepsilon_0}  \sum^{\infty}_{n=1} \frac{1}{R^{n+1}}
\int_W r^n P_n( \cos \theta)  \rho({\bf r}) dV_D. 
\ee
The nth term
\be
V_n=  \frac{1}{4 \pi \varepsilon_0}  \frac{1}{R^{n+1}}
\int_W r^n P_n( \cos \theta)  \rho({\bf r}) dV_D. 
\ee
is usually named according to the following:
n - multipole, 0 - monopole, 1 - dipole, 2 - quadrupole.

\section{Electric Dipole Moment of Fractal Distribution of Charges}
    
An electric multipole expansion is a determination 
of the voltage $V$ due to a collection of charges obtained 
by performing a multipole expansion. 
This corresponds to a series expansion of the charge density 
$\rho({\bf r})$ in terms of its moments, normalized by the distance 
to a point ${\bf R}$ far from the charge distribution. 
In MKS, the electric multipole expansion is given by Eq. (\ref{11}):
\be 
V=\frac{1}{4 \pi \varepsilon_0}\sum^{\infty}_{n=0} 
\frac{1}{R^{n+1}} \int_W r^n P_n(\cos \theta) \rho({\bf r}) dV_D  ,
\ee
where $P_n(\cos \theta)$ is a Legendre polynomial. 

The first term arises from $P_0(\xi)=1$, while all further terms 
vanish as a result of $P_n(\xi)$ being a polynomial in $\xi$ for $n\ge 1$, 
giving $P_n(0)=0$ for all $n\ge 1$.

Set up the coordinate system so that $\theta$ measures 
the angle from the charge-charge line with the midpoint 
of this line being the origin. Then the $n = 1$ term is given by 
\[ V_1= \frac{1}{4 \pi \varepsilon_0} \frac{1}{R^2} \int_W
r P_1(\cos \theta) \rho({\bf r}) dV_D = \]
\[ =\frac{1}{4 \pi \varepsilon_0  R^2} \int_W
r \ \cos \theta \rho({\bf r}) dV_D =
\frac{1}{4 \pi \varepsilon_0 R^2} \int_W
\frac{({\bf r},{\bf R})}{R}  \rho({\bf r}) dV_D = \]
\be =\frac{1}{4 \pi \varepsilon_0 R^3} \int_W
 \left({\bf r},{\bf R}\right) \rho({\bf r}) dV_D =
\frac{1}{4 \pi \varepsilon_0 R^3} 
\left( \int_W {\bf r}  \rho({\bf r}) dV_D , \ {\bf R} \right) .\ee

For a continuous charge distribution, 
the electric dipole moment is given by 
\be \label{d1} {\bf p}^{(3)}=\int_W {\bf r} \rho({\bf r}) dV_3, \ee
where ${\bf r}$ points from positive to negative. 
Defining the dipole moment for the fractal distribution by
the equation
\be \label{ED} {\bf p}^{(D)}=\int_W  {\bf r} \rho({\bf r})  dV_D , \ee
then gives 
\be V_1=\frac{1}{4 \pi \varepsilon_0} 
\frac{({\bf p}^{(D)},{\bf R})}{R^3} =
\frac{1}{4 \pi \varepsilon_0} \frac{p^{(D)} \cos \alpha}{R^2}, \ee
where we use 
\be \cos \alpha=({\bf p}^{(D)},{\bf R})/(p^{(D)} R) , \quad 
p^{(D)}=\sqrt{(p^{(D)}_x)^2+(p^{(D)}_y)^2+(p^{(D)}_z)^2} . \ee

Let us consider the example of electric dipole moment
for the homogeneous ($\rho({\bf r})=\rho_0$) fractal distribution
of electric charges in the parallelepiped region W:
\be \label{paral} 
 0 \le x \le A,\quad  0 \le y \le B , \quad 0 \le z \le C  . \ee
In the case of Riemann-Liouville fractional integral, 
we have $p^{(D)}_x$ in the form
\be p^{(D)}_x=\frac{\rho_0}{\Gamma^3(a)} 
\int^A_0 dx \int^B_0 dy \int^C_0 dz \  x^{a}y^{a-1}z^{a-1}= 
\frac{\rho_0 (ABC)^a}{\Gamma^3(a)}  \frac{A}{a^2(a+1)} ,\ee
where $a=D/3$. 
The electric charge of parallelepiped region (\ref{paral})
is defined by
\be Q(W)=\rho_0 \int_W dV_D=
\frac{\rho_0 (ABC)^{a}}{a^3 \Gamma^3(a)} . \ee
Therefore, we have the dipole moments for fractal
distribution in parallelepiped in the form
\be p^{(D)}_x=\frac{a}{a+1} Q(W) A ,\ee
By analogy with these equation, we can derive
\be p^{(D)}_y=\frac{a}{a+1} Q(W) B, \quad 
p^{(D)}_z=\frac{a}{a+1} Q(W) C . \ee
Using $a/(a+1)=D/(D+3)$, we get
\be  p^{(D)}=\frac{2 D}{D+3} p^{(3)} , \ee
where $p^{(3)}=|{\bf p}^{(3)}|$ are the dipole moment for the usual 
3-dimensional homogeneous distribution. For example, 
the relation $2\le D \le 3$ leads us to the following inequality
\be 0.8 \le p^{(D)}/ p^{(3)} \le 1 . \ee

\section{Electric Quadrupole Moment of Fractal Distribution of Charges}

There are also higher-order terms in 
the multipole expansion that become smaller as $R$ becomes large.
The electric quadrupole term in MKS is given by 
\[ V_2= \frac{1}{4 \pi \varepsilon_0} \frac{1}{R^3} \int_W
r^2 P_2(\cos \theta) \rho({\bf r}) dV_D = \]
\[ = \frac{1}{4 \pi \varepsilon_0} \frac{1}{R^3} \int_W
r^2 (\frac{3}{2}\cos^2 \theta-\frac{1}{2}) \rho({\bf r}) dV_D = \]
\be = \frac{1}{4 \pi \varepsilon_0} \frac{1}{2 R^3} \int_W
( \frac{3}{R^2}({\bf R},{\bf r})^2-r^2) \rho({\bf r}) dV_D .\ee
    
The electric quadrupole is the third term in an electric 
multipole expansion, and can be defined in MKS by
\be V_2= \frac{1}{4 \pi \varepsilon_0}\frac{1}{2 R^3}
\sum^3_{k,l=1} \frac{X_k X_l}{R^2} Q_{kl} , \ee
where $\varepsilon_0$ is the permittivity of free space, 
$R$ is the distance from the fractal distribution of charges, 
and $Q_{kl}$  is the electric quadrupole moment, which is a tensor.
Note that $X_k$ are Cartesian's coordinates of the vector ${\bf R}$,
and $x_k$ are coordinates of the vector ${\bf r}$.

The electric quadrupole moment is defined by the equation
\be \label{EQM} 
Q_{kl}=\int_W [3 x_k x_l-r^2\delta_{kl}] \rho({\bf r}) dV_D ,\ee
where $x_k= x, y$, or $z$. From this definition, it follows that
\be Q_{kl}=Q_{lk} , \quad and  \quad \sum^{3}_{k=1} Q_{kk}=0. \ee
Therefore, we have $Q_{zz}=-Q_{xx}-Q_{yy}$.
In order to compute the values
\be Q^{(D)}_{xx}=
\int_W [2x^2-y^2-z^2] \rho({\bf r}) dV_D , \ee 
\be Q^{(D)}_{yy}=
\int_W [-x^2+2y^2-z^2] \rho({\bf r}) dV_D , \ee 
\be Q^{(D)}_{zz}=
\int_W [-x^2-y^2+2z^2] \rho({\bf r}) dV_D , \ee 
we consider the following expression
\be \label{Qabc}
Q(\alpha,\beta,\gamma)=\int_W [\alpha x^2+\beta y^2+\gamma z^2] 
\rho({\bf r}) dV_D .\ee
Using Eq. (\ref{Qabc}), we have
\be \label{QQ} Q^{(D)}_{xx}=Q(2,-1,-1), \quad  Q^{(D)}_{xx}=Q(-1,2,-1),
\quad Q^{(D)}_{zz}=Q(-1,-1,2) . \ee
The example of electric quadrupole moment for 
the parallelepiped and  ellipsoid regions
are considered in Secs. 6 and 7.

\section{Quadrupole Moment of Charged Fractal Parallelepiped}

Let us consider the example of electric quadrupole moment
for the homogeneous ($\rho({\bf r})=\rho_0$) fractal distribution
of electric charges in the parallelepiped region (\ref{paral}).
If we consider the region $W$ in the form (\ref{paral}), 
then we get
\be \label{Qabc2} 
Q(\alpha,\beta,\gamma)= \frac{\rho_0 (ABC)^a}{(a+2)a^2 \Gamma^3(a) }
[\alpha A^2+\beta B^2+\gamma C^2] , \ee
where we use the Riemann-Liouville fractional integral \cite{SKM}, 
and the function $C_3(D,{\bf r})$ in the form
\be C_3(D,{\bf r})=
\frac{|x y z |^{a-1}}{\Gamma^3(a)} , \quad a=D/3. \ee

The electric charge of the region $W$ is 
\be \label{Qpar}
Q(W)=\rho_0 \int_W dV_D=\frac{\rho_0 (ABC)^a}{a^3 \Gamma^3(a)}  .\ee
If $D=3$, we have $Q(W)=\rho_0 ABC$.
Using equations (\ref{Qabc2}) and (\ref{Qpar}), 
we get the following equation
\be Q(\alpha,\beta,\gamma)= \frac{a}{a+2} Q(W)
[\alpha A^2+\beta B^2+\gamma C^2] . \ee
If $D=3$, then we have $a/(a+2)=1/3$. 
As the result, we have electric quadrupole moments $Q^{(D)}_{kk}$ of 
fractal distribution in the region $W$:
\be Q^{(D)}_{kk}=\frac{3D}{D+6} \ Q^{(3)}_{kk} , \ee
where $Q^{(3)}_{kk}$ are moments for the usual 
homogeneous distribution ($D=3$). 
By analogy with these equations, we can derive $Q^{(D)}_{kl}$ for the case
$k\not=l$. These electric quadrupole moments are
\be Q^{(D)}_{kl}=\frac{4 D^2}{(D+3)^2} \ Q^{(3)}_{kl} , \quad (k\not=l). \ee
Using inequality $2<D< 3$, we get the relations for diagonal elements
\be 0.75  < Q^{(D)}_{kk}/ Q^{(3)}_{kk} \le 1 , \ee
and nondiagonal elements
\be 0.64  < Q^{(D)}_{kl}/ Q^{(3)}_{kl} \le 1 . \ee
where $k\not=l$.

\section{Quadrupole Moment of Charged Fractal Ellipsoid}

Let us consider the example of electric quadrupole moment
for the homogeneous ($\rho({\bf r})=\rho_0$) fractal distribution
in the ellipsoid region $W$:
\be\label{ell} 
\frac{x^2}{A^2}+\frac{y^2}{B^2}+\frac{z^2}{C^2} \le 1 . \ee
If we consider the region $W$ in the form (\ref{ell}), then we get
expression (\ref{Qabc}) in the form
\be Q(\alpha,\beta,\gamma)= \frac{8 \rho_0 (ABC)^a}{ (3a+2) \Gamma^3(a) }
[\alpha A^2 Z_1(a)+\beta B^2 Z_2(a)+\gamma C^2 Z_3 (a) ] , \ee
where $a=D/3$, and $Z_i(a)$, $i=1,2,3$  are defined by
\be Z_1(a)= S(a+1,a-1) S(a-1,2a+1), \ee
\be Z_2(a)= S(a-1,a+1) S(a-1,2a+1),  \ee
\be Z_3(a)= S(a-1,a-1) S(a+1,2a-1) . \ee
Here we use the following function
\be S(n,m)=\int^{\pi/2}_0 dx \ \cos^n (x) \sin^m (x) =
\frac{\Gamma(n/2+1/2) \Gamma(m/2+1/2)}{2 \Gamma(n/2+m/2+1)} . \ee
Note that $Z_1(a)=Z_2(a)=Z_3(a)$.
Using these equations, we get the following relation: 
\be Q(\alpha,\beta,\gamma)= \frac{2 \rho_0 (ABC)^a }{ (3a+2)} 
\frac{\Gamma^2(a/2) \Gamma(a/2+1)}{\Gamma^3(a) \Gamma(3a/2+1)}
[\alpha A^2 +\beta B^2 +\gamma C^2 ] \ee
Using $\Gamma(\beta+1)=\beta \Gamma(\beta)$, we have
\be \label{Qabce}
Q(\alpha,\beta,\gamma)= \frac{2 \rho_0 (ABC)^a}{ 3(3a+2)} 
\frac{\Gamma^3(a/2)}{\Gamma^3(a) \Gamma(3a/2)}
[\alpha A^2 +\beta B^2 +\gamma C^2 ] . \ee

If $D=3$, we obtain
\be 
Q(\alpha,\beta,\gamma)= \frac{4 \pi \rho_0 ABC}{15}
[\alpha A^2 +\beta B^2 +\gamma C^2 ] . \ee
The total charge of the ellipsoid region $W$ is defined by
\be \label{Qe} Q(W)=\rho_0 \int_W dV_D=
\rho_0 (ABC)^a \frac{2 \Gamma^3(a/2)}{3a \Gamma^3(a)\Gamma(3a/2)} .
\ee
If $D=3$, we have the total charge
$Q(W)=(4 \pi/3) \rho_0 ABC $. Here we use $\Gamma(1/2)=\sqrt{\pi}$. 

Using Eqs. (\ref{Qabce}) and (\ref{Qe}), we can derive the electric
quadrupole moments (\ref{QQ}) for fractal ellipsoid.
As the result, we have
\be Q(\alpha,\beta,\gamma)= \frac{D}{3D+6} Q(W)
[\alpha A^2 +\beta B^2 +\gamma C^2 ] . \ee
If $D=3$, then we have the well-known relation:
\be Q(\alpha,\beta,\gamma)= (1/5) Q(W)
[\alpha A^2+\beta B^2+\gamma C^2] . \ee
If $2<D<3$, then we have
\be 
\frac{5}{6}  < Q^{(D)}_{kk} /Q^{(3)}_{kk} < 1 .
\ee

The nondiagonal elements of electric quadrupole moment are 
defined by the following equations
\be 
Q_{xy}=3 \rho_0 \int_W x y \ dV_D , \quad 
Q_{xz}=3 \rho_0 \int_W x z \ dV_D, \quad
Q_{yz}=3 \rho_0 \int_W y z \ dV_D .
\ee
Using these equations, we can derive
the nondiagonal elements in the form:
\be 
Q^{(D)}_{xy}=\frac{6\rho_0 (ABC)^a}{3a+2} 
\frac{\Gamma(a/2) \Gamma^2(a/2+1/2)}{\Gamma^3(a) \Gamma(3a/2+1)} AB,
\ee
\be 
Q^{(D)}_{xz}=\frac{6\rho_0 (ABC)^a}{3a+2} 
\frac{\Gamma(a/2) \Gamma^2(a/2+1/2)}{\Gamma^3(a) \Gamma(3a/2+1)} AC,
\ee
\be 
Q^{(D)}_{yz}=\frac{6\rho_0 (ABC)^a}{3a+2} 
\frac{\Gamma(a/2) \Gamma^2(a/2+1/2)}{\Gamma^3(a) \Gamma(3a/2+1)} BC,
\ee
Using $\Gamma(\beta+1)=\beta\Gamma(\beta)$ and
Eq. (\ref{Qe}), we get the following equations
\be 
Q^{(D)}_{xy}=\frac{6 Q(W)}{3a+2} 
\frac{\Gamma^2(a/2+1/2)}{\Gamma^2(a/2)} AB,
\ee
\be 
Q^{(D)}_{xz}=\frac{6 Q(W)}{3a+2} 
\frac{\Gamma^2(a/2+1/2)}{\Gamma^2(a/2)} AC,
\ee
\be 
Q^{(D)}_{yz}=\frac{6 Q(W)}{3a+2} 
\frac{\Gamma^2(a/2+1/2)}{\Gamma^2(a/2)} BC,
\ee
As the result, we have
\be 
Q^{(D)}_{kl}=\frac{5 \pi}{D+2} 
\frac{\Gamma^2(D/6+1/2)}{\Gamma^2(D/6)} Q^{(3)}_{kl} ,
\ee
where $k\not=l$. Here we use $\Gamma(1/2)=\sqrt{\pi}$.
If we consider $2<D<3$, then we get
\be 0.6972 < Q^{(D)}_{kl}/Q^{(3)}_{kl} < 1 . \ee

\section{Conclusion}

In this paper, we use fractional continuous model for
fractal distribution of electric charges. 
The fractional continuous models for fractal distribution of particles
can have a wide application. 
This is due in part to the relatively small number of parameters 
that define a fractal distribution of great complexity
and rich structure.
In many cases, the real fractal structure of matter can be disregarded 
and the distribution of particles can be replaced by  
some fractional continuous model \cite{PLA05,AP05}. 
In order to describe the distribution with 
noninteger dimension, we must use the fractional calculus.
Smoothing of the microscopic characteristics over the 
physically infinitesimal volume transforms the initial 
fractal distribution into fractional continuous model
that uses the fractional integrals. 
The order of fractional integral is equal 
to the fractal dimension of the distribution.
The fractional continuous model for the fractal distribution
allows us to describe dynamics of
a wide class fractal media \cite{AP05,Chaos05,Physica}.  
One of the dynamical equation of physics is a Liouville equation.
Note that the Liouville equation is a cornerstone of 
the statistical mechanics.  
The fractional generalization of the Liouville equation 
was suggested in Refs. \cite{chaos,JPCS}.
The fractional generalization of the Liouville equation 
allows us to derive the fractional generalization of 
the Bogoliubov equations \cite{PRE05}.
Using fractional analog of the Liouville equation \cite{chaos} and
Bogoliubov equations \cite{PRE05,JPCS}, we can derive the description 
of a fractal distribution as a fractional system. 
Fractional systems can be considered as a special case of 
non-Hamiltonian systems \cite{chaos,PRE05}. 
Note that non-Hamiltonian systems can have
stationary states of the Hamiltonian systems
\cite{Tar-mplb,IJMPB,JPA05,JPA05-2,AP05a,Tarpre}.



\end{document}